\documentclass[%
 reprint,
 amsmath,amssymb,
 aps,
]{revtex4-2}

\usepackage{graphicx}
\usepackage{dcolumn}
\usepackage{bm}
\usepackage{xcolor}
\newcommand{\ket}[1]{| {#1} \rangle} 

\begin{document}

\preprint{APS/123-QED}

\title{Simulating indefinite causal order with Rindler observers}

\author{Aleksandra Dimi{\' c}}
\address{Faculty of Physics, University of Belgrade, Studentski Trg 12-16, 11000 Belgrade, Serbia}
\email{aleksandra.dimic@ff.bg.ac.rs}

\author{Marko Milivojevi{\' c}}
\affiliation{Faculty of Physics, University of Belgrade, Studentski Trg 12-16, 11000 Belgrade, Serbia}

\author{Dragoljub Go{\v c}anin}
\affiliation{Faculty of Physics, University of Belgrade, Studentski Trg 12-16, 11000 Belgrade, Serbia}

\author{Nat\'alia S. M\'oller}

\affiliation{Departamento de F\'isica, Universidade Federal de Minas Gerais, Av. Ant\^onio Carlos, Belo Horizonte, MG, 31270-901, Brazil and
Physics Department and Research Center Optimas, Technische Universit\"at Kaiserslautern, 67663 Kaiserslautern, Germany}

\author{{\v C}aslav Brukner}
\affiliation{Vienna Center for Quantum Science and Technology (VCQ), University of Vienna,
Faculty of Physics, Boltzmanngasse 5, A-1090 Vienna, Austria}
\affiliation{Institute for Quantum Optics and Quantum Information (IQOQI),
Austrian Academy of Sciences, Boltzmanngasse 3, A-1090 Vienna, Austria}

\date{\today}

\begin{abstract}
Realization of indefinite causal order (ICO), a theoretical possibility that even causal relations between physical events can be subjected to quantum superposition, apart from its general significance for the fundamental physics research, would also enable quantum information processing that outperforms protocols in which the underlying causal structure is definite. In this paper, we start with a proposition that an observer in a state of quantum superposition of being at two different relative distances from the event horizon of a black hole, effectively resides in ICO space-time generated by the black hole.   
By invoking the fact that the near-horizon geometry of a Schwarzschild black hole is that of a Rindler space-time, we propose a way to simulate an observer in ICO space-time by a Rindler observer in a state of superposition of having two different proper accelerations. By extension, a pair of Rindler observers with entangled proper accelerations simulates a pair of entangled ICO observers. Moreover, these Rindler-systems might have a plausible experimental realization by means of optomechanical resonators. 

\end{abstract} 

\maketitle

\section{Introduction}

The principle of causality is an implicit assumption of every physical theory and it is universally supported by our experience of nature. From an operational point of view, causality can be understood as a system of signaling or communication relations between physical systems; an information flow whose properties are intimately related to the nature of space and time. One may even say that \emph{the very essence of the classical structure of space and time is to impose a physical constraint on information processing}.

In the old Newtonian picture of the World, space and time are two generically different entities, universal for all observers. There is a single, three-dimensional flat Euclidean space and a single global time that enable us to unambiguously distinguish between past, present and future. Together, they constitute an absolute, independent background structure relative to which every physical event takes place. Signals can propagate in space with unlimited speed (action at a distance) and, consequently, each event can be caused by any other in its present or past. The special theory of relativity (SR) changed this paradigm: space and time became united into the $(3+1)$-dimensional \emph{space-time continuum} - Minkowski space - in which signals cannot travel faster than the speed of light, enforcing them to stay within the local light cone. Nevertheless, the structure of Minkowski space adhered {to} the character of an independent, fixed background on which dynamical matter fields propagate.

The radical change came with Einstein's general theory of relativity (GR). {The} gravitational field came to be understood as the curved space-time itself, gravity being encoded in a metric tensor coupled to dynamical matter fields. There is no fixed, independent metric structure, no absolute background stage relative to which locations of physical events are to be defined, there are just dynamical fields, {the} metric being one of them, and physical events can only be located \emph{relative to each other}. The possibility of communication between different observers, i.e. the causal order, is entirely determined by the dynamical configuration of light cones, and so, although dynamical, space-time, as a landscape of physical events, has definite causal order (DCO). Thinking about Quantum Mechanics (QM) of gravitating objects, 
{the} question arises whether there is a way to relax the restrictions of classical space-time structure and enable processes that do not obey definite causal relations, i.e. can there be a quantum superposition of different causal orders - an \emph{indefinite causal order} (ICO)?

It is generally expected that unification of QM and gravitational physics will provide us with some deeper insights concerning the nature of space and time and their relationship with matter. 
However, the standard methods of quantization of matter fields employed in Quantum Field Theory (QFT) do not seem to work for Einstein's gravity; it holds a status of a non-renormalizable effective theory with undetermined high-energy degrees of freedom. In order to surpass the traditional concepts of GR and QFT, various ways of ``quantizing'' gravity were proposed so far, such as String Theory, Quantum Loop Gravity, Noncommutative Geometry, Supergravity, etc. However, to date, there has been no conclusive empirical evidence that would support or disprove any of the proposed ``high-energy theories''. This state of affairs motivates us to reconsider in which sense and to what extent can the seemingly contradictory principles of QM and GR be reconciled, while adhering  {to} the tenets of both theories \cite{Hardy1, Hardy2}.


There are two main incentives for this paper. The first came from the work of Oreshkov, Costa and Brukner \cite{OCB12}, where it was found that it is possible to formulate quantum mechanics without any reference to a global causal structure, i.e. without predefined space-time. The resulting framework - the \emph{process matrix formalism} - allows for processes incompatible with any definite causal order between operations performed on quantum systems. These abstract indefinite causal structures are shown to be advantageous for quantum computing \cite{CDP+13,ACB14} and quantum communication \cite{FAB15,GFA+16,C17}. One particular example that has {an} experimental demonstration is the so called ``quantum switch'' \cite{CDP+13, C12, FDD+14,PMA+15,RAK+16,RRF+17,GCK+18}, where the main idea is to use an auxiliary quantum system that can coherently control the order in which certain operations are applied. In the case of the so-called gravitational quantum switch (GQS) \cite{ZCP+ar} the role of the control system is played by a gravitating object prepared in a state of quantum superposition of being at two different spatial locations. 
The second incentive comes from the intriguing idea of \emph{quantum reference frames} (QRF) \cite{Giac17, Felipe18} where one regards reference frames not as abstract systems of coordinates, but as actual physical objects subjected to the laws of quantum mechanics and describes the world from their perspective. 

In this paper, we propose a way to simulate ICO processes by considering the fact that the near-horizon geometry of a Schwarzschild black hole (BH) is that of a Rindler space-time. 
Namely, a Rindler observer in a state of superposition of having two different proper accelerations corresponds to a near-horizon Schwarzschild observer in a state of superposition of being at two different locations along a single radial direction.  From the viewpoint of such an observer the geometry of space-time is indefinite. This correspondence can be extended to a pair of Rindler observers with entangled proper accelerations simulating a pair of entangled ICO observers, as we illustrate by means of a simple example. Although they represent idealizations, these Rindler-systems could become a valuable resource for studying ICO processes in laboratory conditions, with plausible experimental realization in the form of opto-mechanical oscillators \cite{ASP15,KBS15,ADS+16,EMR}.


\section{Rindler observers}
In order to set the stage, consider the (1+1)-dimensional Minkowski space $\mathcal{M}_{2}$ and a central light cone defined by $t=\pm x$ (we set $c=1$). In these globally inertial coordinates $(t,x)$ the Minkowski metric is given by $ds^{2}_{\mathcal{M}_{2}}=-dt^{2}+dx^{2}$. If we introduce \emph{Rindler coordinates} $(\eta,\rho)$ defined by $t=\rho\sinh(\eta)$ and $x=\rho\cosh(\eta)$, the metric becomes $ds^{2}_{\mathcal{M}_{2}}=-\rho^{2}d\eta^{2}+d\rho^{2}$. However, these new coordinates do not cover the whole Minkowski space, only the patch given by $x\geq 0$ and $\vert t\vert\leq x$. This region is called the \textit{right Rindler wedge} or simply the R-wedge (see Fig. 1 (left panel)). The family of coordinate lines of constant $\rho$ are the branches of hyperbolas $x^{2}-t^{2}=\rho^{2}$ embedded in the R-wedge (the other set of branches belongs to the L-wedge, defined by $x\leq0$ and $\vert t\vert\leq -x$) asymptotically approaching the Rindler horizon $t=\pm x$ ($\rho=0$ and $\eta\rightarrow\pm\infty$). They correspond to the worldlines of physical systems that have constant proper acceleration of magnitude $\alpha(\rho)=1/\rho$ - \emph{Rindler observers}. The proper time of a Rindler observer, with a given $\rho=const.$, is $d\tau=\rho d\eta$. Therefore, we can define a worldline of a Rindler observer in the R-wedge with proper acceleration $\alpha$ by a pair of parametric equations:
\begin{equation}\label{eq:0}
  t(\tau)=\frac{1}{\alpha}\sinh(\alpha\tau),\;\; x(\tau)=\frac{1}{\alpha}\cosh(\alpha\tau).
\end{equation}
Rindler observer with greater proper acceleration has {a} more curved worldline (closer to the Rindler horizon). Note also that, due to the presence of the Rindler horizon, Rindler observers in the R-wedge are causally disconnected from the ones in the L-wedge, meaning that they are unable to communicate with each other.

Consider now a pair of Rindler observers in the R-wedge, with different proper accelerations $\alpha_{1}$ and $\alpha_{2}$. Let the worldline of the second one be more curved. That is, let $\alpha_{1}<\alpha_{2}$. A photon sent to the left from the source $S$, located at $t_{s}=0$ and $x_{s}=x_{0}>0$, intersects worldlines of the  {Rindler observers} at proper times $\tau_{1}$ and $\tau_{2}$, respectively (see Fig. 1 (right panel)). At $t=0$ both observers are closer to the origin than $S$, implying that $\alpha_{2}x_{0}>\alpha_{1}x_{0}>1$. This configuration has an interesting feature that will turn out to be important. Namely, given the values of $x_0$ and $\alpha_{1}$, there exists \emph{a unique} value of $\alpha_{2}$, defined as the non trivial solution ($\alpha_2\neq\alpha_1$) of the equation
\begin{equation}\label{eq::16}
  \alpha_2 x_0= (\alpha_1 x_0)^{\frac{\alpha_2}{\alpha_1}},
\end{equation}
for which $\tau_{1}=\tau_{2}$ (for details, see Appendix A).  
\begin{figure}
\center
\includegraphics[width=9cm]{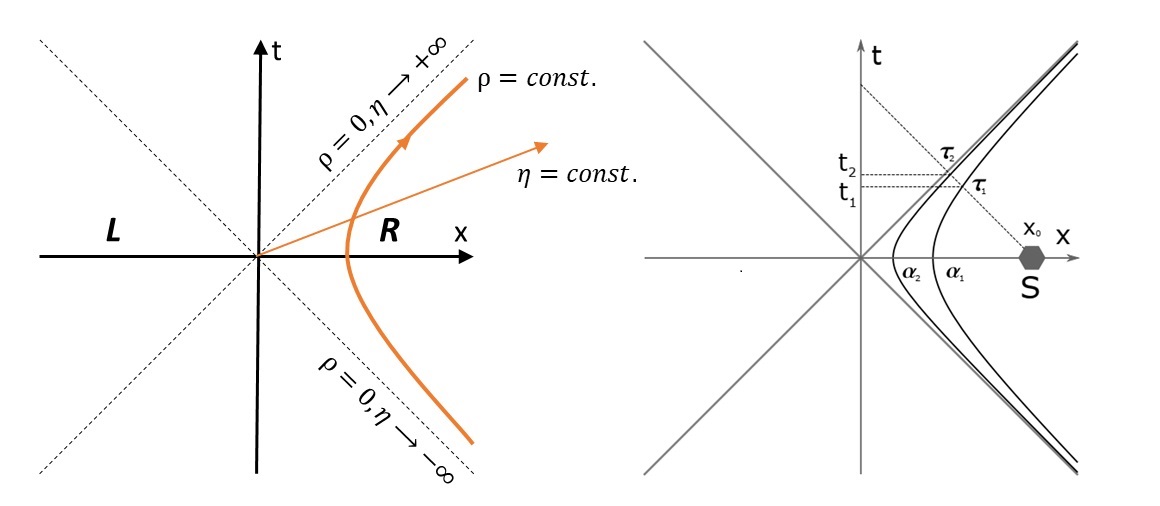}\label{lin}
\caption{{\it Rindler observers.} (Left panel): {\emph{Coordinate lines in the right (R) Rindler wedge of the Minkowski space.}} Coordinate lines of constant $\rho$ are hyperbolas and they correspond to worldlines of Rindler observers. Regions L and R, called the left and the right Rindler wedge, respectively, are causally disconnected due to the presence of the Rindler horizon $t=\pm x$. (Right panel): {\emph{Photon's worldline intersects the Rindlers.}} A pair of Rindler observers in the R-wedge with different proper accelerations $\alpha_{1}$ and $\alpha_{2}$, $\alpha_{1}<\alpha_{2}$. A photon sent from the source $S$, located at $t_{s}=0$ and $x_{s}=x_{0}>0$, intersects the worldlines of the  {Rindler observers} at their respective proper times $\tau_{1}$ and $\tau_{2}$. One can arrange things so that $\tau_{1}=\tau_{2}$.}
\end{figure}
\section{Indefinite causal order via Rindler observers}
Let us assume that we have a pair of Rindler observers in the R-wedge, Rindler-Amber ($A_{R}$) and Rindler-Blue ($B_{R}$). { \emph{Amber} and \emph{Blue} are the colors by which we distinguish the two observers}, see Fig. 2. Note, however, that these ``observers'' need not be actual macroscopic measuring devices of any sort, nor sentient beings; they could be microscopic physical systems with some internal degrees of freedom (like spin). On the other hand, they have definite worldlines since they are confined within accelerating laboratories (imagine well-enough localized classical ``boxes'' each carrying an atom). We assume that these internal degrees of freedom are such that they do not get affected by the accelerated motion of the Rindler laboratory, which we also assume to be completely isolated. Source $S$ emits a photon whose worldline intersects the worldlines of $A_{R}$ and $B_{R}$. The photon starts in some polarization state $\ket{\Psi}$ and the {Rindler observers} can perform instantaneous unitary transformations on it.

When their worldlines intersect, $A_{R}$ performs a unitary transformation $U_{A}$ on the photon's polarization state. This constitutes event $a$. In general, $U_{A}$ is a function of $A_{R}$'s proper time. 
We can abstractly think of $A_{R}$ as a Rindler ``clock'' whose worldline and the ticking rate are defined by the proper acceleration $\alpha_{A}$. The state of $A_{R}$ will therefore be denoted by $\ket{\tau_{\alpha_{A}};A}$, without getting into details of what $A_{R}$'s actual degrees of freedom are and what kind of Hamiltonian governs the dynamics thereof.  
And the same protocol applies to $B_{R}$. Its meeting with the photon and application of a unitary transformation $U_{B}$ constitutes event $b$. By choosing a suitable values of the proper accelerations $\alpha_{A}$ and $\alpha_{B}$, we can arrange that meetings of the {Rindler observers} with the photon (events $a$ and $b$) occur \emph{at the same proper times} $\tau_{a}=\tau_{b}=\tau^{*}$ (see the discussion at the end of Section $2$). 


Here we want to stress that physical events are not regarded as pure geometrical points that constitute space-time manifold (modulo diffeomorphisms) with some definite set of causal relations defined by the metric. Rather they are defined operationally, through application of a specific unitary transformation, or more generally a specific  {completely positive trace-preserving} (CPTP) map. Taking quantum mechanics into account, we consider the possibility that the same physical event can be in a superposition of occurring at different space-time locations. This would enable the realization of indefinite causal order between pairs of events, such as $a$ and $b$ in the above discussion. One example of this situation is the already mentioned quantum switch \cite{C12,FDD+14,PMA+15,RAK+16,RRF+17,GCK+18}.

On the left panel of Fig. $2$, the proper acceleration of $A_{R}$ ($\alpha_{2}$) is greater than the proper acceleration of $B_{R}$ ($\alpha_{1}$) and on the right panel, the values are interchanged, $A_{R}$ has the smaller proper acceleration ($\alpha_{1}$) and $B_{R}$ has the greater proper acceleration ($\alpha_{2}$).
In the reference frame of the inertial observer sitting at $x=0$ the initial state (at $t=0$) of the whole system (Rindlers $\otimes$ photon) in the former case is the separable state $\ket{\tau_{\alpha_2}(0),A}\ket{\tau_{\alpha_1}(0),B}\ket{\Psi}$; the photon first meets $B_{R}$ and then $A_{R}$. In the latter case, when the initial state of the system is $\ket{\tau_{\alpha_1}(0),A}\ket{\tau_{\alpha_2}(0),B}\ket{\Psi}$, the photon first meets $A_{R}$ and then $B_{R}$. If the {Rindler observers} are prepared in the \textit{entangled state} that is a superposition of the two previous ones, at $t=0$ we have
\begin{align}\label{eq:bpofp}
\frac{1}{\sqrt{2}}\Big(&\ket{\tau_{\alpha_{2}}(0),A}\ket{\tau_{\alpha_1}(0),B} \nonumber\\
+&\ket{\tau_{\alpha_1}(0),A}\ket{\tau_{\alpha_2}(0),B}\Big)\ket{\Psi}.
\end{align}

It is important to realize that the event $a$ has to be \emph{one and the same} in both ``branches'' of the superposition. The meeting of the photon with $A_{R}$ in the case when $A_{R}$ has greater proper acceleration than $B_{R}$ and the meeting of the photon with $A_{R}$ when $A_{R}$ has smaller proper acceleration than $B_{R}$, have to be locally indistinguishable events in every respect. That is why we demand that the event occurs at the same proper time, $\tau^{*}$, in both situations. {In principle, the state of the photon could be affected by the kinematic state of a Rinldler laboratory, which might give rise to entanglement between proper accelerations of the Rindler laboratories and photon's polarization state. To avoid this possibility, we put them to rest just before they meet the photon, thus making them inertial from that point on. Conditional deacceleration of the laboratories along the Rindler trajectories can be performed sufficiently fast, yet gradually, not to produce the Unruh radiation.}


For $t<t_{1}$ (where $t_{1}$ is the time coordinate of the intersection of the photon's worldline with the less curved Rindler worldline) the state is
\begin{align}\label{eq:before}
\frac{1}{\sqrt{2}}\Big(&\ket{\tau_{\alpha_{2}}(t),A}\ket{\tau_{\alpha_1}(t),B} \nonumber\\
+ &\ket{\tau_{\alpha_1}(t),A}\ket{\tau_{\alpha_2}(t),B}\Big)\ket{\Psi}.
\end{align}
{By the time the photon went through the laboratories, unitary transformations, $U_{A}(\tau^{*})$ and $U_{B}(\tau^{*})$, have been applied on it.} At some instant $t>t_{2}$ (where $t_{2}$ is the time coordinate of the intersection of the photon's worldline with the more curved Rindler worldline) the state of the whole system is given by
\begin{align}\label{eq:afterpofp}
\frac{1}{\sqrt{2}}\Big(&\ket{\tau^{*}+t-t_2,{A}}\ket{\tau^{*}+t-t_1,{B}}U_{A}(\tau^{*})U_{B}(\tau^{*}) \\
+&\ket{\tau^{*}+t-t_1,{A}}\ket{\tau^{*}+t-t_2, {B}}U_{B}(\tau^{*})U_{A}(\tau^{*})\Big)\ket{\Psi},  \nonumber
\end{align}
where $t-t_1$ and $t-t_2$ are the time intervals during which the respective Rindler laboratories were at rest. 

Finally, we need to disentangle the state of the photon from the state of the {Rindler observers}. To this end, at some moment $t_m$, a projective measurement (postselection {on} the internal state of the Rindlers) is performed in the superposition basis $\{\ket{m_{i}}, \ket{m_{i}^{\perp}}\vert\;\; i=1,2\}$, separately for each laboratory. The basis states are given by
\begin{align}\label{eqmtau}
\ket{m_{i}}&=\frac{1}{\sqrt{2}}\Big(\ket{\tau^{*}+t_m-t_i, A}+\ket{\tau^{*}+t_m-t_i, B} \Big), \nonumber \\
\ket{{m_{i}^{\perp}}}&=\frac{1}{\sqrt{2}}\Big(\ket{\tau^{*}+t_m-t_i, A}-\ket{\tau^{*}+t_m-t_i, B}\Big).
\end{align}
Postselection on any pair of possible measurement results leads to the final state of the photon
\begin{equation}\label{final_photon_state}
\frac{1}{\sqrt{2}}(U_{A}(\tau^{*})U_{B}(\tau^{*})\pm U_{B}(\tau^{*})U_{A}(\tau^{*}))\ket{\Psi}.
\end{equation}
 
Subsequently, the photon may be observed at $C$. However, being in the state (\ref{final_photon_state}), there is no way to distinguish, given the photon alone, which of the two events ($a$ and $b$) lies in the causal future of the other, {and the information about the causal order is lost.}

\begin{figure}[t]
\center
\includegraphics[width=8cm]{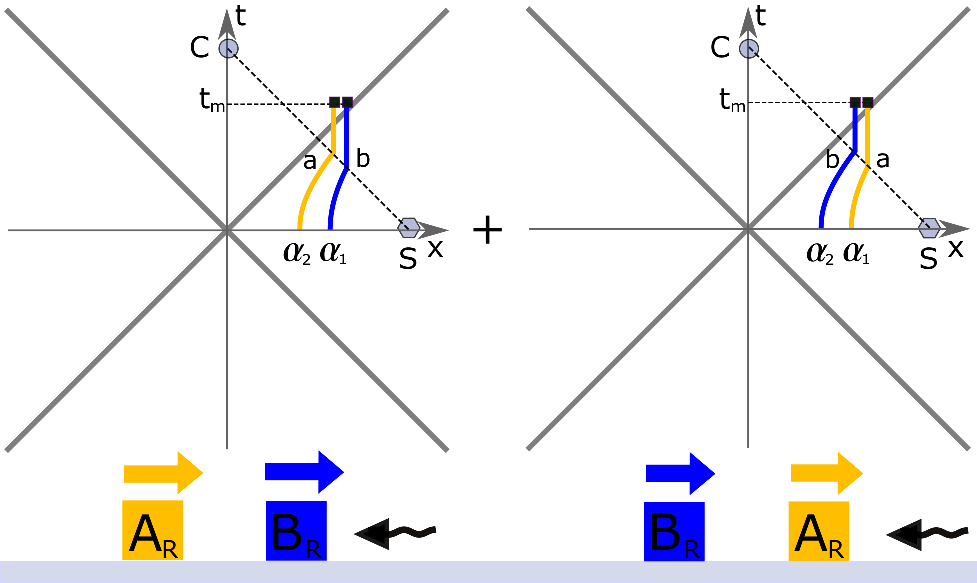}\label{switch}
\caption{{\it Entangled Rindler observers.} Worldline of a photon emitted by the source $S$ intersects worldlines of the two Rindler observers, $A_{R}$ and $B_{R}$, that have entangled proper accelerations. On the left panel, $A_{R}$ has greater proper acceleration ($\alpha_{2}$) than $B_{R}$ ($\alpha_{1}$), and on the right panel the accelerations are ``switched''. By choosing a suitable pair of values for the proper accelerations, these meetings (events $a$ and $b$) occur at the same proper time $\tau^{*}$. Rindler observers act on the polarization of the photon according to their ``color'', amber ($A$) or blue ($B$), that distinguishes them. Observer $A_{R}$ performs a unitary transformation $U_{A}$ at $\tau^{*}$ and observer $B_{R}$ performs a unitary transformation $U_{B}$ at $\tau^{*}$. When {Rindler observers} meet the photon, they come to rest and remain that way until some particular moment $t_{m}$ at which a projective measurement is performed in order to disentangle the state of the photon from that of the {Rindler observers}. The photon is observed at $C$.}
\end{figure}

\section{Gravitational scenario}
Imagine now that we have a system that involves a Schwarzschild BH 
and an observer (outside the horizon) in a state of superposition of being at two different relative proper distances from the horizon. The observer is well-enough localized and has a negligible effect on the gravitational field. Also, we do not assume the existence of a fixed background geometry with reference to which we could define positions; only the \emph{relative} distance between the BH (its horizon) and the observer has physical meaning.


In Schwarzchild coordinates $(t,r,\theta,\phi)$, the metric of the BH exterior is 
\begin{equation}
ds^{2}=-f(r)dt^{2}+\frac{dr^{2}}{f(r)}+r^{2}d\Omega^{2}_{2},
\end{equation}
with $f(r)=1-\frac{R_{S}}{r}$ and $\Omega_{2}^{2}=d\theta^{2}+\sin^{2}{\theta}d\phi^{2}$. We are only interested in a single radial direction, so we can ignore the angular part of the metric. The proper radial distance between the stationary observer at $r_{lab}$ and the event horizon at $R_{S}=2MG$ ($M$ being the mass of the BH) is 
\begin{equation}
\rho=\int_{R_{S}}^{r_{lab}}\frac{dr}{\sqrt{f(r)}}.
\end{equation}

Therefore, our bipartite system can be interpreted as a situation where we have an observer in the state $\tfrac{1}{\sqrt{2}}(\vert\rho_{1}\rangle+\vert\rho_{2}\rangle)$ with indefinite proper distance from the horizon in definite Schwarzchild geometry. On the other hand, from the viewpoint of the observer (quantum reference frame), the gravitational field 
appears to be indefinite, as if the BH is in the state of superposition of being at two different places relative to the observer. 

The idea that a gravitating object in a state of quantum superposition of being at two different locations somehow ``induces'' a quantum superposition of different geometries, dates back to Feynman \cite{Feynman} and it has been successfully promoted recently \cite{ZCP+ar, Ana15, Bose17, Vedral, Rovelli}. Although this seems as a natural way to combine GR and the linearity of QM, it remains unclear in which sense can a gravitational field (i.e. space-time geometry) be in a state of quantum superposition, see for example \cite{Penrose}. Here, we propose a way of looking at this situation based on the relational character of quantum superposition {\cite{ZychCostaRalph}}.

Einstein's equivalence principle states that a gravitational field is \emph{locally equivalent} to an accelerating reference frame in flat space-time. As a consequence, 
for every well-enough localized stationary $(r=\rm{const}.)$ observer in Schwarzschild geometry there is an equivalent uniformly accelerating observer in Minkowski space.
Moreover, the near-horizon geometry of a Schwarzschild BH is that of a Rindler space-time, and therefore a Schwarzschild observer whose proper distance from the horizon is $\rho$ corresponds to the Rindler observer whose proper acceleration is $1/\rho$ (see Appendix B for details). 

By extending this reasoning, we propose a {``quantum''} version of Einstein's equivalence principle by stating that ICO space-time is locally equivalent to a non-inertial reference frame with superposed proper accelerations. In particular, we can relate a Schwarzschild observer in the state $\tfrac{1}{\sqrt{2}}(\vert\rho_{1}\rangle+\vert\rho_{2}\rangle$) of being at two different relative distances from the horizon, to the Rindler observer in the state of superposition of having proper accelerations $1/\rho_{1}$ and $1/\rho_{2}$, respectively. The similar principle has been invoked and derived within quantum reference frames formalism \cite{Giac17} in the Newtonian limit.

Consider now a pair of observers in the near-horizon region of a Schwarzschild BH, Schwarzschild-Amber $(A_{S})$ and Schwarzschild-Blue $(B_{S})$. Let the observers have entangled proper distances from the horizon (along a single radial ray) and fixed relative distance between each other. From the reference frame of the BH, space-time has {a} definite geometry and the state of this tripartite system is $\tfrac{1}{\sqrt{2}}\ket{0}_{BH}(\vert \rho_{1}\rangle_{A}\vert \rho_{2}\rangle_{B}+\vert \rho_{2}\rangle_{A}\vert \rho_{1}\rangle_{B})$, where $\ket{0}_{BH}$ is the position state of the black hole. Observers $A_{S}$ and $B_{S}$ correspond to a pair of Rindler observers, $A_{R}$ and $B_{R}$, both in the R-wedge, with \emph{entangled proper accelerations}. 
On the other hand, from the point of view of the pair of observers, we have a BH in a state of superposition of being ``at two different sides'' of the observers, symmetrically. These states of the BH are denoted by $\ket{L}$ and $\ket{R}$, see Fig. 3. The middle point between $A_S$ and $B_S$ is well defined in relative terms. From the reference frame associated to this point \cite{Varn} the joint state of the system is $\frac{1}{\sqrt{2}}(\ket{L}+\ket{R})\ket{-\bar{\rho}}_A\ket{\bar{\rho}}_B$,
where $\bar{\rho}=\frac{\rho_2-\rho_1}{2}$ is the half-distance between the two 
observers.


We can perform a photon experiment, similar to the one described in Section 3, that involves the observers $A_{S}$ and $B_{S}$ and a source $S$ that emits a photon in the direction that depends on the position of the BH relative to the observers. 
The position of the BH plays the role of a quantum control for the whole process (gravitational quantum switch). Due to the gravitational time dilation, we can arrange things so that the photon passes through both laboratories at the same moment of their local proper time (see Appendix C for details). This is analogous to the case of the Rindler quantum switch from Section $2$. When the photon gets inside the laboratory, a unitary transformation, $U_{A}$ or $U_{B}$, depending on the laboratory, is applied instantaneously on its polarization state. The meeting of the photon and the laboratory $A_{S}$ and the application of the unitary $U_{A}$ is the \emph{event a}, and likewise, the meeting of the photon and the laboratory $B_{S}$ and the application of the unitary $U_{B}$ is the \emph{event b}. After performing a projective measurement in the superposition basis $1/\sqrt{2}(\vert L\rangle\pm \vert R\rangle)$ of the BH (to disentangle its state from the photon's state, as described in \cite{ZCP+ar}), the final state of the photon implies the two events do not possess definite causal order.

\begin{figure}[t]
\center
\includegraphics[width=8cm]{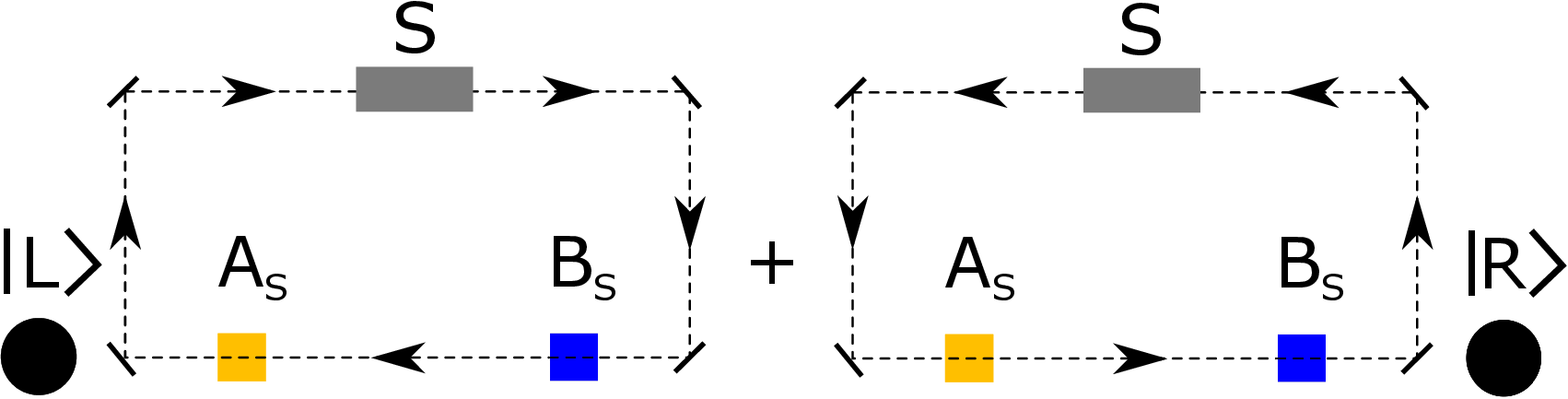}\label{fig3}
\caption{\textit{Gravitational quantum switch}. The system involves a photon source $S$, two observers, $A_{S}$ and $B_{S}$, and a BH in a state of superposition of being at two different positions relative to them, $\vert L\rangle$ or $\vert R\rangle$. The direction in which the photon is emitted depends {on} the position of the BH and therefore the BH plays the role of a quantum control for the whole process. As a result, the photon is in a superposition of traveling in two opposite directions.}
\end{figure}

\section{Conclusion} 
In this paper, we proposed a way to characterize ICO space-time as a space-time associated with the reference frame of a quantum observer - quantum observer perceives ICO space-time. As an illustration, we considered a bipartite system that involves a Schwarzschild BH and an observer outside the horizon, in a state of quantum superposition of being at two different relative distances. By invoking the fact that near-horizon geometry is that of a Rindler space-time, we can relate this ICO observer to the Rindler observer in a state of superposition of having two different proper accelerations. 
By extension, a pair of ICO observers with entangled proper distances from the BH horizon corresponds to a pair of Rindler observers with entangled proper accelerations. As an example, we analyzed Rindler quantum switch and the related gravitational quantum switch.

Furthermore, a Bell's inequality for temporal order of events was found in \cite{ZCP+ar}. The same kind of inequality can be derived by using two pairs of Rindler observers, one in the left and the other in the right Rindler wedge. In the corresponding gravitational scenario we would have to take into account the Kruskal extension of the Schwarzschild solution. In this case, the gravitational quantum switch would involve two pairs of observers residing in conformally flat space-times connected by Einstein-Rosen bridge. We postpone this interesting analysis for future work.


On a more practical side, there is a growing effort in demonstrating quantum features of nano-to-mesoscale optomechanical systems. This may provide a challenging, yet feasible experimental realizations for the proposed Rindler systems \cite{ASP15}. Recently, mesoscopic mechanical resonators were considered as quantum non-inertial reference frames \cite{KBS15,ADS+16} and entanglement of two massive mechanical oscillators is achieved \cite{EMR}. It has been proposed to utilize quantum optical fields in order to prepare and measure the quantum states of mechanical resonators, conceivably opening the possibility to quantum-mechanically control the acceleration of such quantum non-inertial reference frames \cite{ASP15}.

In an actual experiment, potential decoherance effects can compromise the predicted result \cite{Decoh}. Moreover, QFT effects could also be taken into account. In this context, our Rindler observers could be viewed as Unruh-DeWitt detectors, where an increase of the thermal noise, due to the Unruh effect, may affect the evolution of the system, such that it can no {longer} be considered as a coherent superposition, but rather a (convex) classical mixture. However, since we can choose proper accelerations of the Rindler observers to be arbitrarily small by putting the photon source at suitable position, the Unruh effect can always be made negligible.
{Just to put some numbers, if we set $x_0=1 \; \textrm{m}$, the accelerations would be of order $10^{17}\; \textrm{m/s}^2$, which corresponds to the Unruh temperature of order $10^{-4} \; \textrm{K}$, and this is far too small for the Unruh effect to be detectable. Correspondingly, for a solar mass black hole with $R_s = 3\; \mathrm{km}$, we have a Schwarzschild observer at $1 \mathrm {m}$ proper distance from the horizon, which is a good near-horizon approximation.} Furthermore, depending on the parameters of the objects involved, e.g. masses of the laboratories, these Rindler systems could be used to test hypotheses such as the Ghirardi-Rimini-Weber (GRW) model of objective collapse \cite{GRW}. Namely, a failure to maintain a coherent macroscopic superposition even after screening off the system from decoherence effects, might be taken as an indication of a spontaneous GRW-type collapses. However, as we noted in the text, the systems involved in our setting need not be macroscopic systems (they can be microscopic {ones}, such as atoms). Moreover, the issues concerning macroscopic systems may still be avoided by performing the experiment ''sufficiently fast'' (before the alleged GRW-type of collapse should take place). This kind of assessment was provided in \cite{ZCP+ar} for gravitational quantum switch. Finally, we should also mention that there is a relativistic version of the GRW model \cite{Tumulka}. Our relativistic Rindler systems could perhaps be used for studying and testing such models.

Finally, and perhaps most importantly, we would like to establish a more rigorous framework that would allow us to formally define an ICO space-time related to a general quantum reference frame. This could be an important step towards a better understanding of the quantum nature of space-time.

{\bf Acknowledgments}\newline
This is a significantly revised version of the paper that can already be found on the arXiv  \cite{Rindler}. 
The authors thank Dejan Simi\'{c}, Marko Vojinovi\'{c}, Nikola Paunkovi\'{c} and \"{A}min Baumeler for helpful comments. A.D., M.M. and D.G. acknowledge support from the bilateral project SRB 02/2018, \textit{Causality in
Quantum Mechanics and Quantum Gravity}, between Austria and Serbia. A.D. and M.M. acknowledge support from the project No. ON171035 and D.G. from the project No. ON171031 of Serbian Ministry of Education and Science. Additionally, A.D. acknowledges support from scholarship awarded from The Austrian Agency for International Cooperation in Education and Research (OeAD-GmbH). A.D. and D.G. acknowledge grant FQXi-MGA-1806 that supported their stay in Vienna. A.D. and D.G. would also like to thank University of Vienna and IQOQI for hospitality during their stay. 
N.S.M. thanks Marcelo Terra Cunha, Nelson Yokomizo, Raphael Drumond, Leonardo Neves, and Davi Barros for useful discussions. N.S.M. acknowledge Coordena\c{c}\~ao de Aperfei\c{c}oamento de Pessoal de N\'ivel Superior (CAPES), Conselho Nacional de Desenvolvimento Cient\'ifico e Tecnol\'ogico (CNPq) and Funda\c{c}\~ao de Amparo \`a Pesquisa do Estado de Minas Gerais (FAPEMIG) for financial support and VI Paraty Quantum Information School for the introduction into the topic.
{\v{C}}.B. acknowledges the support from the research platform TURIS, from the European Commission via Testing the Large-Scale Limit of Quantum Mechanics (TEQ) (No. 766900) project, and from the Austrian-Serbian bilateral scientific cooperation no. 451-03-02141/2017-09/02, and by the Austrian Science Fund (FWF) through the SFB project  BeyondC and a grant from the Foundational Questions Institute (FQXi) Fund. This publication was made possible through the support of the ID 61466 grant from the John Templeton Foundation, as part of the The Quantum Information Structure of Space-time (QISS) Project (qiss.fr). The opinions expressed in this publication are those of the  author(s) and do not necessarily reflect the views of the John Templeton Foundation. \\[0.25cm]

{\bf Author Contribution.} A.D., M.M., D.G. and \v{C}.B conceived the research. A.D., M.M. and D.G. provided main theoretical idea and equally contributed to writing the paper. N.M. participated in discussion concerning entangled laboratories. {\v C}.B. supervised the project.\\[0.12cm]

\appendix
\setcounter{figure}{0}
\renewcommand{\thefigure}{A\arabic{figure}}
\section*{APPENDIX}
\section{Equal proper times condition for a pair of Rindler observers}

Here we give a simple derivation of the relation (\ref{eq::16}) that has to be satisfied if a photon's worldline is to intersect a pair of Rindler worldlines at the same moment of their proper times. We set $c=1$.

From (Fig. A1) we can see that
\begin{equation}\label{eqA::1}
x_0-x_{i}=t_{i},
\end{equation}
where $i\in\{1,2\}$. {From the parametric equations of the Rindler worldlines,
\begin{equation}\label{eqA::3}
   t_{i}=\frac{1}{\alpha_{i}}\sinh(\alpha_{i}\tau_{i}),\;\;x_{i}=\frac{1}{\alpha_{i}}\cosh({\alpha_{i}\tau_{i}}),
\end{equation}
we can deduce the instant of the observer's proper time $\tau_{i}$
in which he/she receives the signal from the source $S$ sitting at $x_0>0$. }From (\ref{eqA::1}) and (\ref{eqA::3}) we get
\begin{equation}\label{eqA::5}
  \frac{{\rm e}^{\alpha_{i}\tau_{i}}+{\rm e}^{-\alpha_{i}\tau_{i}}}{2\alpha_{i}}+\frac{{\rm e}^{\alpha_{i}\tau_{i}}-{\rm e}^{-\alpha_{i}\tau_{i}}}{2\alpha_{i}}=x_{0},
\end{equation}
%
\begin{figure}[htp]
\centering
\includegraphics[width=5.5cm]{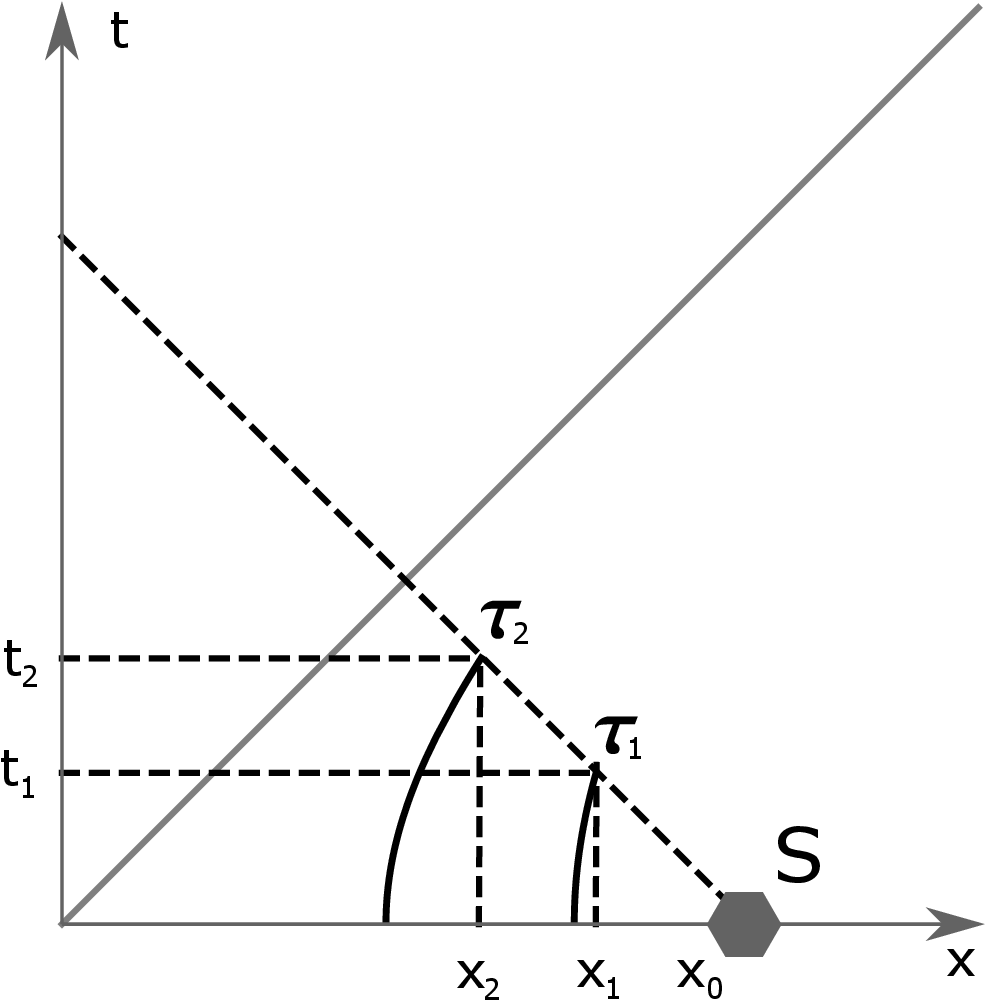}\label{lin1}
\caption{{\it A photon intersecting two Rindler observers.}
Worldline of a photon sent from the source $S$ intersects worldlines of two Rindlers at instants $\tau_{1}$ and $\tau_{2}$ of their respective proper time. Global inertial coordinates of the intersection points are denoted by $t_i$, $x_i$, $i\in\{1,2\}$.}
\end{figure}
%
and so,
\begin{equation}\label{eqA::6}
 \tau_{i}=\frac{1}{\alpha_{i}}\ln(\alpha_{i} x_{0}).
\end{equation}
\noindent
Note that both the prefactor and the argument of the logarithm are positive.
The equality condition for the two proper times, $\tau_{1}$ and $\tau_{2}$, gives us the following relation between $\alpha_{1}$ and $\alpha_{2}$
\begin{equation}\label{eqA::16}
  \alpha_{2}x_0= (\alpha_{1}x_0)^{\frac{\alpha_{2}}{\alpha_{1}}}.
\end{equation}
By introducing new variables,
$X:=\alpha_{1}x_0$ and $Y:=\alpha_{2}x_0$, previous equation can be formulated as
\begin{equation} \label{eqA::17}
  Y=X^{\frac{Y}{X}}.
\end{equation}
Numerical analysis of (\ref{eqA::17}) shows that {the} solution $Y=X$, which exists for each value of $X$, is unique for $X<1$
(Fig. A2 (a)).
This solution is trivial, since it corresponds to a single Rindler observer (or a pair of overlapping Rindler observers with $\alpha_{1}=\alpha_{2}$).
If $X>1$, there are two possible solutions for $Y$, the trivial one ($Y=X$) and the other that can be greater than $X$, if $X\in(1,e)$, or less than $X$, if $X>e$ (Fig.
A2 (b,c)). Note that for every value of $x_{0}$ (position of the source) we have a continuous infinity of pairs of Rindler trajectories that satisfy (\ref{eqA::16}).

\begin{figure}[h!]
\center
\includegraphics[width=8.7cm]{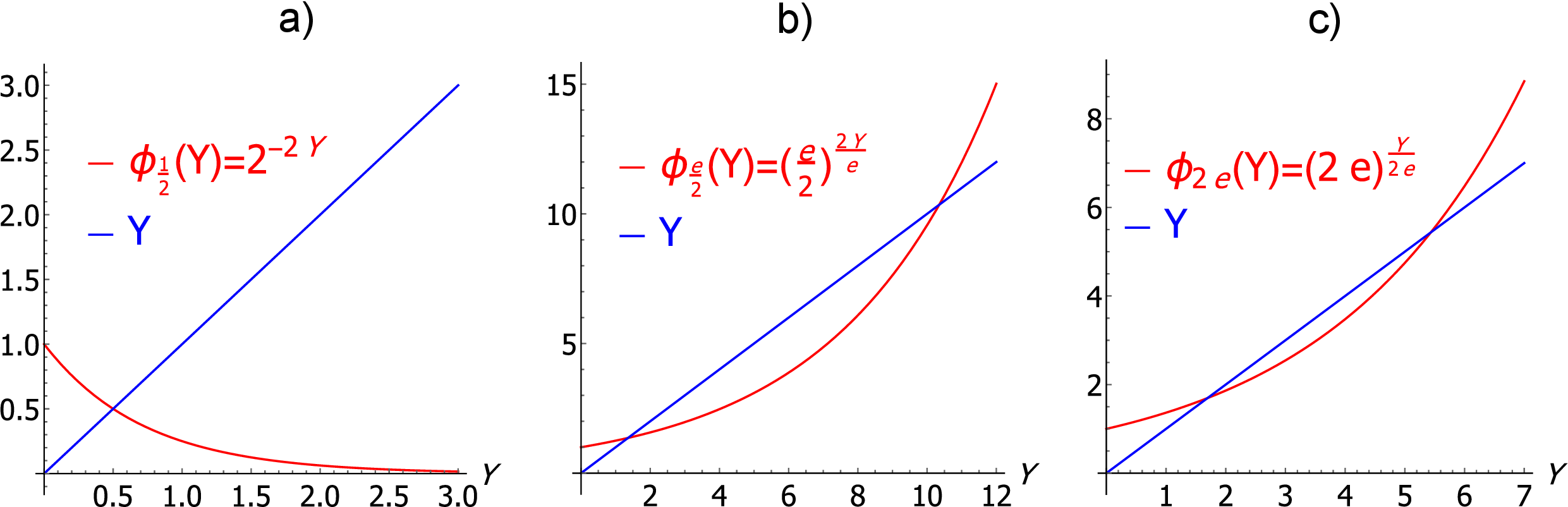}\label{lin2}
\caption{{\it Numerical analysis. Acceleration ratio for a pair of Rindler observers.} Setting $X:= \alpha_{1}x_0$ and $Y:=\alpha_{2}x_0$ equation (\ref{eqA::16}) becomes $Y=\Phi_X(Y)=X^{\frac{Y}{X}}$. {By plotting the left and right hand side of $Y=\Phi_X(Y)$ as functions of $Y$, we find the intersection points for a given value of $X$, which plays the role of a parameter for the family of functions $\Phi_{X}(Y)$. Trivial solution is given by $Y=X$, and it corresponds to an uninteresting case of equal accelerations. So, without loss of generality, we assume $Y>X$ (i.e., $\alpha_{2}>\alpha_{1}$). We examine three different case, depending on the range of values of the parameter $X$. Diagram a) represents $X<1$ case, in particular $X=1/2$, and here we only have one intersection corresponding to the trivial solution. Diagram b) represents $X \in (1,{\rm e})$ case, in particular $X={\rm e}/2$, and there we have two intersection, the second one corresponding to a non-trivial solution for which $Y>X$. 
Finally, in the case $X>{\rm e}$, we again have two solutions - the trivial one, and the other for which
$Y<X$. That case is represented in c), with $X=2e$.}}
\end{figure}

\setcounter{figure}{0}
\renewcommand{\thefigure}{B\arabic{figure}}

\section{Near-horizon geometry of a Schwarzschild black hole}
Worldlines of stationary observers in \emph{curved space-time} do not conform to  {the} geodesics defined by the space-time metric. To maintain a fixed position they must oppose their inertia with some proper acceleration. Generally, {the} $4$-acceleration of an observer in {a} curved spacetime is given by $a^{\mu}=U^{\nu}\nabla_{\nu}U^{\mu}$, where $U^{\mu}$ is its $4$-velocity, and its proper acceleration by $\alpha=\sqrt{g_{\mu\nu}a^{\mu}a^{\nu}}$. In particular, for a stationary observer in {the} Schwarzschild metric  
\begin{equation}\label{SCW_metric}
ds^{2}=-f(r)dt^{2}+\frac{dr^{2}}{f(r)}+r^{2}d\Omega^{2}_{2},
\end{equation}
with $f(r)=1-\frac{R_{S}}{r}$ and $\Omega_{2}^{2}=d\theta^{2}+\sin^{2}{\theta}d\phi^{2}$, we have
\begin{equation}
\alpha=\frac{R_{S}}{2r^{2}\sqrt{f(r)}}. \label{aSproper}
\end{equation}
Analogously to the proper time $d\tau=\sqrt{f(r)}dt$ of an observer sitting at $r=\rm{const.}$, we can introduce \emph{proper radial distance},  
\begin{equation}
d\rho=\frac{dr}{\sqrt{f(r)}}.
\end{equation}
Integrating from $R_{S}$ to $r$ we get the proper radial distance of the stationary observer sitting at some fixed $r$ from the event horizon,
\begin{align}
\rho=&r\left(1-\frac{R_{S}}{r}\right)^{1/2}\nonumber\\
&+\frac{R_{S}}{2}\ln\left[\frac{2r}{R_{S}}-1+\frac{2r}{R_{S}}\left(1-\frac{R_{S}}{r}\right)^{1/2}\right]. \label{Connecton}
\end{align}
From (\ref{Connecton}) it follows that $\rho\sim r$ in the limit $r/R_{S}\rightarrow +\infty$, and so from (\ref{aSproper}) we get $\alpha\sim R_{S}/2\rho^{2}$, which is just the Newtonian inverse square law. 
Now, let $r=R_{S}+\epsilon$ for some small $\epsilon$. In the limit $\epsilon/R_{S}\rightarrow 0$ we have $\rho\sim 2 \sqrt{R_{S}\epsilon}$, and the proper acceleration of a stationary observer in the vicinity of the event horizon is inversely proportional to its proper distance from the horizon, $\alpha\sim 1/\rho$. In the intermediate region, proper acceleration is some complicated function of the observers proper radial distance.

{The} Schwarzschild metric (\ref{SCW_metric}) near the horizon ($r\approx R_{S}$) becomes $Rindler\times S^{2}$,
\begin{equation}
ds^{2}=-\rho^{2}d\eta^{2}+d\rho^{2}+R_{S}^{2}d\Omega^{2}_{2}, \label{near horizon metric}
\end{equation}
where we introduced a new time coordinate $\eta:=\frac{t}{2R_{S}}$. The non-angular part of the above metric is the metric of $(1+1)$-dimensional Minkowski space $\mathcal{M}_{2}$ in Rindler coordinates. This is evident if we start with the metric of $\mathcal{M}_{2}$ in Minkowski coordinates, $ds^{2}_{\mathcal{M}_{2}}=-dt^{2}+dx^{2}$,
and introduce Rindler coordinates $(\rho,\eta)$ by $t=\rho\sinh(\eta)$ and $x=\rho\cosh(\eta)$ in which the metric takes the form $ds^{2}_{\mathcal{M}_{2}}=-\rho^{2}d\eta^{2}+d\rho^{2}$.

{Since $x^{2}-t^{2}=\rho^{2}\geq 0$, coordinates $(\rho,\eta)$ cover only the part of $\mathcal{M}_{2}$ defined by $x\geq 0$ and $\vert t\vert\leq x$ - \emph{the right Rindler wedge}.} Rindler coordinates $(\rho,\eta)$ become singular at $\rho=0$ but, using the Minkowski coordinates $(t,x)$, one could analytically continue them from the R-wedge to the whole Minkowski space. Similarly, in the case of a Schwarzschild black hole, we use Kruskal coordinates to make an analytic continuation of {the} Schwarzschild coordinates $(t,r)$ across the horizon thus obtaining their maximal extension. The event horizon of a black hole, defined by $\rho=0$, corresponds to the light cone $t=\pm x$, and the near-horizon black hole geometry is $Rindler\times S^{2}$.
An observer at $r=\rm{const.}$ $(r\approx R_{S})$ in Schwarzschild metric corresponds to a uniformly accelerating observer with $\rho=\rm{const.}$ in the right Rindler wedge, that is, an observer in Minkowski space whose worldline is a hyperbola $x^{2}-t^{2}=\rho^{2}=\rm{const.}$, and whose constant proper acceleration is given by
\begin{equation}
\alpha=\frac{1}{\rho}=\frac{1}{2\sqrt{R_{S}}\sqrt{r-R_{S}}}.
\end{equation}
\setcounter{figure}{0}
\renewcommand{\thefigure}{C\arabic{figure}}
\begin{figure}[h]
\center
\includegraphics[width=2.9cm]{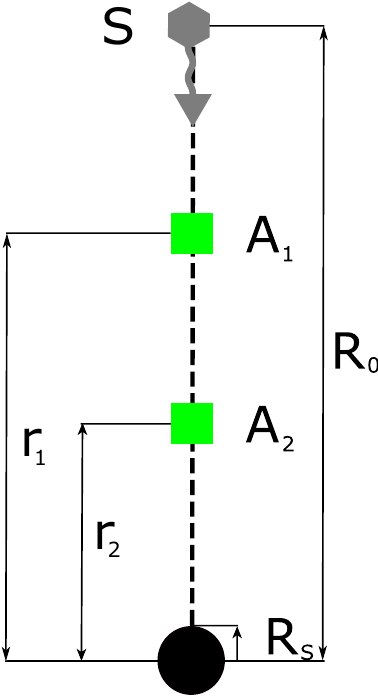}\label{Pad}
\caption{{\it Schematic representation of a BH and two stationary laboratories $A_1$ and $A_2$.} Radially falling photon passes through $A_{1}$ and $A_{2}$ sitting at two different radial distances from the BH. The fact that ticking rates of the local clocks are different (with respect to Schwarzschild time $t$) makes it possible to arranged the relative distances so that meetings of the photon with the laboratories occur at the same moment of their local proper time.}
\end{figure}
\section{Equal proper times for a free falling photon}
Start with the equation for {a} radially free falling photon in Schwarzschild coordinates
\begin{equation}
\frac{dt}{dr}=-\frac{1}{1-\frac{R_S}r}=-\frac{r}{r-R_S},
\end{equation}
where $R_S$ is the Schwarzschild radius.
A radially falling photon, emitted from the source $S$ at $r=R_0$, passes through the laboratory $A_1$, sitting at $r=r_{1}$, at the time $t_{1}$. A simple calculation gives us that for $t_{1}$ we have
\begin{equation}
t_{1}=R_0-r_{1}+R_S\ln{\frac{R_0-R_S}{r_{1}-R_S}}.
\end{equation}
The same photon passes through the laboratory $A_2$ sitting at $r=r_{2}$, at the time $t_{2}$ given by (see Fig. C1)
\begin{equation}
t_{2}=R_0-r_{2}+R_S\ln{\frac{R_0-R_S}{r_{2}-R_S}}.
\end{equation}
We are looking for the condition for both events to happen at the same local proper time measured in laboratories $A_1$ and $A_2$, that is
\begin{equation}\label{equaltau}
\tau_{1}=\tau_{2}.
\end{equation}
Proper times of the laboratories are related to the global time $t$ (proper time of a stationary observer at infinity) by
\begin{equation}
\tau_{i}=\sqrt{1-\frac{R_S}{r_{i}}}t_{i}.
\end{equation}
{After inserting (C2) and (C3) into (C5), simple numerical analysis tells us that it is not possible to
obtain the relation (\ref{equaltau}) for arbitrary values of $R_0$ and $R_S$.
However, if $R_S/R_0>10^{-4}$ (the photon source cannot be arbitrarily far away), there is a continuum of pairs of different $r_{1}$ and $r_{2}$ that satisfy the condition of equal proper times.}

\end{document}